\documentclass[10pt]{article}
\usepackage{ulem}
\usepackage{epsfig,graphicx}
\usepackage{amsmath}
\usepackage{amssymb}
\usepackage{array}
\usepackage{slashed}
\usepackage{color}
\newcommand{\beq}{\begin{equation}}
\newcommand{\eeq}{\end{equation}}
\newcommand{\bea}{\begin{eqnarray}}
\newcommand{\eea}{\end{eqnarray}}

\newcommand{\OMIT}[1]{{}}

\newcommand\spur{\raise.15ex\hbox{/}\kern-.57em }

\newcommand{\lsim}{
\mathrel{\hbox{\rlap{\hbox{\lower4pt\hbox{$\sim$}}}\hbox{$<$}}}}
\newcommand{\gsim}{
\mathrel{\hbox{\rlap{\hbox{\lower4pt\hbox{$\sim$}}}\hbox{$>$}}}}

\newcommand{\calL}{{\cal{L}}}
\newcommand{\calS}{{\cal{S}}}
\newcommand{\calO}{{\cal{O}}}
\newcommand{\barg}{{\bar{g}}}
\newcommand{\barA}{{\bar{A}}}
\newcommand{\barD}{{\bar{D}}}
\newcommand{\tilA}{{\tilde{A}}}
\newcommand{\tilD}{{\tilde{D}}}
\newcommand{\IH}{{\rm{I\!H}}}

\newcommand{\eff}{{\rm{eff}}}

\newcommand{\bfk}{{\bf{k}}}

\newcommand{\rr}{{\bf r}}
\newcommand{\q}[2]{ {\bf q}_{#1}^{#2}}

\newcommand{\setq}[2]{\{{\bf q}_{#1}^{#2}\}}

\begin{document}

\begin{flushright}
LA-UR-11-00458
\end{flushright}
\vspace{2.0 true cm}
\begin{center}
{\LARGE {\bf
A low energy theory for superfluid and solid matter and its application to the neutron star crust
}}\\
\date{\today}
\vspace{2.0 true cm}
{\large Vincenzo Cirigliano$^a$,  Sanjay Reddy$^a$,  Rishi Sharma$^{a,b}$}\\
\vspace{1.0 true cm}
${}^a$
{\sl Theoretical Division, Los Alamos National Laboratory, Los Alamos, NM 87545, USA} \\
\vspace{0.2 true cm}
${}^b$ {\sl  Theory Group, TRIUMF, 4004 Wesbrook Mall, Vancouver, BC, V6T 2A3, Canada} \\
\vspace{0.2 true cm}
\today
\end{center}
\vspace{1.5cm}

\begin{abstract}
We formulate a low energy effective theory describing phases of matter that are both solid and
superfluid. These systems simultaneously break translational symmetry and the phase symmetry associated with 
particle number. The symmetries restrict the combinations of terms that can appear in the effective action
and the lowest order terms featuring equal number of derivatives and Goldstone fields are completely
specified by the thermodynamic free energy, or equivalently by the long-wavelength limit of static
correlation functions in the ground state. We show that the underlying interaction between particles
that constitute the lattice and the superfluid gives rise to entrainment, and mixing between the
Goldstone modes.   As a concrete example we discuss the low energy theory for the inner crust of a
neutron star, where a lattice of ionized nuclei coexists with a neutron superfluid.
\end{abstract}

\newpage
\section{Introduction}
The low energy dynamics of strongly interacting solids and superfluids can be systematically studied
through an effective theory formulation in terms of weakly interacting phonons - the collective
degrees of freedom in these systems. In the familiar case of solids, one longitudinal phonon and two
transverse phonons arise as Goldstone modes due to the breaking of translation symmetry. In the case
of a superfluid, one mode called the superfluid phonon arises due to the breaking of the global
$U(1)$ symmetry associated with phase rotations of a field operator \footnote{The $U(1)$
symmetry is related to particle number conservation and we will refer to this as a phase symmetry. 
Its breaking simply refers to the choice of a ground state: total number is conserved and the continuity 
equation remains valid. }. In special cases
the ground state of the system can spontaneously break both these symmetries. A particularly simple
but non-trivial realization is a solid immersed in a superfluid with strong interactions between the
particles that form the solid and the superfluid respectively.  It is likely that a substantial
region in the crust of a neutron star is occupied by such a  phase~\cite{Baldo:2005} and its
presence may affect neutron star phenomenology. From general considerations we can argue that the
inner crust of neutron stars features a lattice of neutron rich nuclei in a bath of unbound
superfluid neutrons. The lattice sites can be viewed as clusters of protons, with a fraction of
neutrons ``entrained'' on the clusters~\cite{Carter:2004pp,Carter:2006}.  Other intrinsically more
complex phases where a single component exhibits both superfluid and solid characteristics have also
been proposed. They include the supersolid phase of $^4$He \cite{Andreev:1969} and the Larkin
Ovchinnikov Fulde Ferrell (LOFF) phases~\cite{FF:1964,LO:1965} in polarized fermion superfluids.
Although these systems can in principle be realized terrestrially, they have proven to be
challenging to explore in experiments~\cite{Kim:2003}.  Nonetheless in all these cases the low
energy dynamics is described by an effective theory of four Goldstone modes \cite{Son:2005ak}. The
associated fields for the lattice phonons are $\xi^{a=1..3}({\bf{r}}, t)$ and are related to
space-time dependent deformations of the lattice.  Similarly, the field associated with the
superfluid mode $\phi({\bf{r}}, t)$ is related to the space-time dependent phase of the condensate.
Because of interactions, such as those between the neutrons and the protons in the neutron star
crust, one can not in general treat the two sectors separately and a unified treatment is required.
It is the aim of this paper to provide such a framework.

The low energy theory is described in terms of the fields $\phi$ and $\xi^a$. The symmetries associated with translation and number conservation require that the low energy theory be invariant under the transformation $\xi^{a=1..3}({\bf{r}}, t)\rightarrow \xi^{a=1..3}({\bf{r}}, t) + a^{a=1..3} $ and $\phi({\bf{r}}, t)\rightarrow \phi({\bf{r}}, t) + \theta$ where $a^{a=1..3} $
and $\theta$ are constant shifts. This naturally implies that the low energy lagrangian can contain only spatial and temporal gradients of these fields.  
Further, by requiring cubic symmetry for the crystalline state, the 
quadratic part of the effective lagrangian is given by,
\begin{equation}
\begin{split}
{\cal{L}} = \frac{f_\phi^2}{2}&(\partial_0\phi)^2 - \frac{v_\phi^2f_\phi^2}{2}(\partial_i\phi)^2
+\frac{\rho}{2}\partial_0\xi^a\partial_0\xi^a - \frac{1}{4} {{\mu}}(\xi^{ab}\xi^{ab}) -
\frac{{K}}{2}(\partial_a\xi^a)(\partial_b\xi^b) \\
&-\frac{\alpha}{2}\sum_{a=1..3}(\partial_a\xi^a\partial_a\xi^a) +{g_{\rm
mix}f_\phi\sqrt{\rho}}~\partial_0\phi\partial_a\xi^a + \cdots~\label{eq:phenomLagrang}\;, 
\end{split}
\end{equation}
where higher order terms involve higher powers of the gradients of these fields, and 
$\xi^{ab} = 
(\partial_a\xi^b+\partial_b\xi^a) - \frac{2}{3}\partial_c\xi^c\delta^{ab}$. In the uncoupled case,
the low energy coefficients (LECs) appearing above, such as $\rho,{\mu},{K}$ are related to
the mass density, the shear modulus, and the compressibility of the solid respectively. They
determine the velocities of the phonons in the solid phase.  Similarly, the velocity of the phonon
in the pure superfluid case is given by $v_\phi$. In the presence of strong coupling between the solid
and superfluid these coefficients are modified.  For example, the coefficient $\rho$ in 
Eq.~\ref{eq:phenomLagrang} differs from the usual mass density of the pure lattice component due
interactions that entrain the superfluid, and the mixing coefficient $g_{\rm mix}$ couples superfluid
and lattice dynamics. As we will show Galilean invariance relates $g_{\rm mix}$  to the modifications of $\rho$ and
$v_\phi$ due to entrainment \cite{Pethick:2010}. An analysis
of these modifications in the context of the neutron star crust due to the underlying interaction
between neutrons and protons was the original motivation for this study. In this case, the mixing
coefficient $g_{\rm mix}$ is relevant for heat transport properties in the inner
crust \cite{Aguilera:2008ed}, and the eigenmodes of the coupled superfluid-solid system could play a role in 
explaining the observed quasi-periodic oscillations in magnetars flares~\cite{Strohmayer:2006py}.

We will present a general proof that the functional form of the lowest-order Lagrangian is
completely specified by the thermodynamic pressure in the presence of constant external fields that
couple to the conserved densities and currents in the system. The derivatives of the pressure  with
respect to these external fields determine the low energy constants. Non-perturbative techniques
such as Quantum Monte Carlo or Hartree-Fock techniques may be suited to calculate these
thermodynamic functions. For example, the energy as a function of the density for a
non-relativistic uniform Fermi gas at unitarity was calculated using Quantum Monte Carlo techniques 
in~\cite{Gezerlis:2007fs}. (For a recent example of the calculation of the LECs in relativistic 
superfluids, see~\cite{Anglani:2011cw}.) Since these derivatives of the thermodynamic functions are related to the
long-wavelength limit of static correlation functions of currents and densities, their direct
calculation using non-perturbative methods also provide the needed LECs.

The outline of the paper is as follows. In Section~\ref{Section:Formal_development} we outline the
formalism and define notation. In Section~\ref{Section:Neutron} we revisit the
effective theory of a neutron superfluid in the absence of any lattice, derived
earlier in Ref.~\cite{Son:2002zn,Son:2005rv}.  This serves as a pedagogic
warm-up before describing  the more complicated system including the lattice.
Here, we prove that lowest order lagrangian is determined  by the thermodynamic
pressure. In Section~\ref{Section:Mixed} we derive results for the relevant case
of the combined neutron and proton sectors. In
Section~\ref{Section:Application} we focus on the applications of our formalism
to the neutron star crust and use simple estimates for the mixing coefficient in
the neutron star inner crust to determine the resulting eigenmodes of the
longitudinal lattice and the superfluid phonons. Here we also comment on
the connection with previous work on the elastic properties of LOFF
phases~\cite{Mannarelli:2007}. We present our conclusions in
Section~\ref{Section:conclusion}.
\section{Symmetries of the underlying Hamiltonian ~\label{Section:Formal_development}}
The prototypical system we consider here is composed of two conserved species of particles. In the
neutron stars,  this would be the strongly coupled many-body system of non-relativistic
neutrons and protons. In what follows we will continue to refer to these two components as neutron
and protons but it should be understood that our considerations apply more generally.  We represent
the action for the system abstractly as ${\cal{S}}[\Psi_n, \Psi_p]$ where $\Psi_n$ and $\Psi_p$, are
the neutron and proton fields. Even though the particles may be non-relativistic, we will work in a
Lorentz covariant form and take the non-relativistic limit at the end of the calculation. We will
denote the spatial indices by Latin characters ($a$, $b$, $c$, $i$, $j$, $k$) running from $1$ to
$3$ and the full space-time indices by
Greek characters ($\mu$, $\nu$, $\sigma$, $\lambda$) running from $0$ to $3$.

The theory describing the neutrons and protons is invariant with respect
to global phase rotations of the neutron field,
$\Psi_n(x)\rightarrow\exp(-i\theta_n) \Psi_n(x)$. The conservation law
associated with the symmetry is the conservation of neutron number. Similarly, independent phase
rotations of the protons gives rise to the conservation of proton number. We note that 
protons and neutrons are separately conserved on timescales small compared to the weak interaction
time $\tau_{\rm weak}\simeq 1/(G_F^2~n_B~T^2)$ where $n_B$ and $T$ are the baryon density and
temperature of the system.  For typical conditions the timescale for low energy dynamics  is $\tau_{\rm
EFT} \simeq 1/T$ and correspondingly 
the ratio $\tau_{\rm EFT} / \tau_{\rm weak} \simeq G_F^2 n_B T \ll 1$. We shall refer to the
corresponding conserved currents as $j^\mu_n$ and $j^\mu_p$ respectively. The action $\calS$ is also
invariant under space-time translations and spatial rotations and the conserved current associated
with translations is the stress energy tensor $T^{\mu\nu}$.

To analyze the constraints provided by these symmetries on the low energy
effective action, it is useful to introduce external fields that couple to the
conserved currents (see e.g. discussion in Ref.~\cite{Leutwyler:1993gf}).
For the internal symmetries we add to the
action  ${\cal{S}}[\Psi_n, \Psi_p]$ source terms of the form  $\int d^4x\;
j^\mu_n(x)A_\mu^n(x)$ and $\int d^4x\; j^\mu_p(x)A_\mu^p(x)$. If the external
fields are allowed to transform appropriately under local $U(1)$ transformations, this
procedure promotes the global symmetries to local symmetries.  This is equivalent to converting all partial
derivatives $\partial_\mu$ to covariant derivatives $\partial_\mu-iA_\mu$. 

A similar extension of space-time symmetries is slightly more subtle. It requires extending
space-time from a flat space-time to a curved space-time, and writing the action in a form that is
general coordinate invariant. Indeed, the non-relativistic theory of neutrons and protons may be seen as
the non-relativistic, flat-space limit, of a fully relativistic, general-coordinate-invariant action.
The external field that couples to the stress-energy tensor is a deformation of the metric, $\delta g_{\mu\nu}$. 

Therefore, we start with an action of the form,
${\cal{S}}[\Psi_n,\Psi_p,A_\mu^n,A_\mu^p,g_{\mu\nu}]$
which is invariant under general coordinate transformations,
\begin{equation} 
\begin{split} 
x^\mu  & \rightarrow  x^{'\mu} = x^\mu + a^\mu (x) \\ 
g^{\mu \nu} (x) & \to g^{'\mu \nu}  (x') =   
g^{\rho\sigma}(x)\frac{\partial x^{'\mu}}{\partial x^\rho }
\frac{\partial x^{'\nu} }{\partial x^\sigma}\\ 
A^\mu(x)&\rightarrow A^{'\mu}(x')=A^{\sigma}(x)
\frac{\partial x^{'\mu} }{\partial x^\sigma}\;,\label{eq:gr1} 
\end{split} 
\end{equation}
these transformations include both rotations and boosts as special cases of local space and time translations.
The action is also invariant under local phase rotations of neutrons,
\begin{equation}
\begin{split}
\Psi_n(x)&\rightarrow\Psi_n'(x)=\exp(-i\theta_n(x))\Psi_n(x)\\
A^n_\mu(x)&\rightarrow
A^{'n}_{\mu}(x)=A^n_{\mu}(x)-\partial_\mu\theta^n(x)\label{eq:gauge_n}\;,
\end{split}
\end{equation}
and local phase rotations of protons,
\begin{equation}
\begin{split}
\Psi_p(x)&\rightarrow\Psi'_p(x)=\exp(-i\theta^p(x))\Psi_p(x)\\
A^p_\mu(x)&\rightarrow
A^{'p}_{\mu}(x)=A^p_{\mu}(x)-\partial_\mu\theta^p(x)\label{eq:gauge_I}\;.
\end{split}
\end{equation}

The reason for extending the global symmetries to local symmetries is that
correlation functions of the conserved currents can be analyzed very simply by
taking appropriate functional derivatives of the generating functional
$W[A_\mu^n,A_\mu^p,g_{\mu\nu}]$ 
with respect to the external fields $A_\mu^n(x), A_\mu^p(x)$ and $g_{\mu \nu} (x)$. The generating
functional is defined in the standard path integral representation as 
\begin{equation}
\begin{split}
e^{iW[A_\mu^n,A_\mu^p,g_{\mu\nu}]}&= \int  [d \Psi_n][d \Psi_p] 
e^{i {\cal S} [\Psi_n, \Psi_p, A_\mu^n, A_\mu^p, g_{\mu\nu}]}\\
& =  Z[A_\mu^n,A_\nu^p,g_{\mu\nu}] \,;
\end{split}
\end{equation}
and  $Z[A_\mu^n,A_\nu^p,g_{\mu\nu}]$ is thermodynamic partition function. For example, the derivative
with respect to the zeroth component of the external field $A_\mu$ defines the number density as given by  
\begin{equation}
\begin{split}
\langle \Omega| \hat{n}_n(x) | \Omega\rangle_{A_\mu^n,A_\mu^p,g_{\mu\nu}} 
&=\frac{\delta W[A_\mu^n, A_\mu^p, g_{\mu\nu}]}{\delta A_0^n(x)}
=\frac{1}{i  Z}  \frac{\delta Z[A_{\mu}^n, A_{\mu}^p, g_{\mu\nu}]}{\delta A^n_0(x)} ~.
\label{eq:number}
\end{split}
\end{equation}

In order to evaluate correlation functions in an equilibrium state with specified number density,
the functional derivatives with respect to $A_\mu(x)$ are evaluated at specific values corresponding to 
appropriate chemical potentials as required by Eq.~\ref{eq:number}.
For neutrons $A^n_\mu(x)=\barA^{n}_{\mu}=(\mu_n+m_n,{\bf{0}})$, where $\mu_n$ is the usual
non-relativistic chemical potential and similarly for protons
$A^p_\mu(x)=\barA^{p}_{\mu}=(\mu_p+m_p,{\bf{0}})$ and $\mu_p$ is the corresponding non-relativistic
chemical potential.
Moreover,  functional derivatives
with respect to $g_{\mu \nu} (x)$ are evaluated at space-time metric $g_{\mu \nu} (x)=\barg_{\mu
\nu}$. For the pure neutron sector (Section~\ref{Section:Neutron}), it suffices to set the equilibrium
metric to be the Minkowski metric, $\barg_{\mu\nu}=\eta_{\mu\nu}$. In the case of coupled system, since 
the spatial components of a space-time independent metric specifies the lattice structure, we will allow $\barg_{\mu\nu}$
to be more general in Section~\ref{Section:Mixed}. 
In the following we discuss specific cases in which the ground state
$|\Omega\rangle$ spontaneously breaks number and translation symmetries of the underlying Hamiltonian.
First, in Section \ref{Section:Neutron} we discuss the simple case of a superfluid which breaks the $U(1)$ symmetry associated with number conservation, and subsequently in Section \ref{Section:Mixed} we discuss the system of interest where both the global $U(1)$ and space-time translation symmetries are simultaneously broken. 
%
\section{One component superfluid~\label{Section:Neutron}}
\subsection{Fields and the effective lagrangian} 
To illustrate the main ideas we first consider a single component superfluid such as degenerate neutron matter where attractive interactions lead to the formation of Cooper pairs and a transition to a superfluid state.
Here, the two-neutron operator has a non-zero expectation value and in equilibrium 
\begin{equation}
\langle\Omega|\Psi_n(x)\Psi_n^T(x)|\Omega\rangle=C\gamma^5\Theta(x)
=C\gamma^5|\Theta|\label{eq:eqm_condensate}\;.
\end{equation}
Since the phase of the condensate changes on making a global phase rotation on
$\Psi_n$, the condensate spontaneously breaks 
$U(1)_n$ to $Z_2$. 
The corresponding Goldstone boson field $\phi(x)$ is given by  the phase fluctuations of the order
parameter. 
 
The field  $\phi(x)$ transforms nonlinearly  under $U(1)_n$  transformations 
with parameter $\theta_n$: 
\begin{equation}
\phi  \to \phi + \theta_n ~. 
\end{equation}
 
The effective Lagrangian for $\phi$ can be in principle obtained by integrating out the heavy modes
corresponding to the gapped fermionic excitations from the system in a Wilsonian approach.   
 
The partition function in  presence of external gauge fields admits the following low-energy
representation: 
\begin{equation}
Z_n[A_\mu^{n}, g_{\mu\nu}] =
\int  [d {\Psi}_n]  e^{i {\cal S} [{\Psi}_n,  A_\mu^{n}, g_{\mu\nu}]} 
\  \longrightarrow \ 
\int [d\phi] 
e^{i {\cal S}_{\rm eff}  [\partial_\mu\phi, A_\mu^{n}, g_{\mu\nu}]}\;\label{eq:ZnLE}.
\end{equation}
The symmetries of the underlying fundamental theory impose stringent constraints on the form of 
the effective lagrangian defined by 
${\cal S}_{\rm eff} =\int d^4x  \sqrt{-g} \, \calL_{\rm{eff}}$.
Global  $U(1)_n$ symmetry implies that the Goldstone boson can 
occur only through the derivative  $\partial_\mu\phi$ and local $U(1)_n$ symmetry (Eq.~\ref{eq:gauge_n}) 
implies that $A_\mu^{n}$ and $\phi$ can appear in $\calL_{\rm{eff}}$ only in the combination 
\begin{equation}
 D_\mu\phi(x)=\partial_\mu\phi(x)+A_\mu^{n}(x)\;.
 \end{equation}
Since we are working in a covariant theory, $\calL_{\eff}$ should transform as
a scalar density  under general coordinate transformations. Therefore, the effective lagrangian 
can only be constructed from building blocks
like $D_\mu\phi$, $D_\nu D_\mu \phi $
etc., with all indices contracted.

In this work we use the power counting scheme proposed by Son and Wingate~\cite{Son:2005rv} to organize terms
in $\calL_{\eff}$. We define the power of an operator as the difference between the
number of $\partial$'s and the power of $\phi$. I.e. all terms of order $(\partial)^m(\phi)^n$
($m\geq n$) with the same value of $m-n$ are considered to have the same order, $m-n$. $A_\mu^{n}$
has the same order as $\partial_\mu\phi$, i.e. order $0$, and $g_{\mu\nu}$ also has order
$0$. Therefore, 
\begin{equation}
\begin{split}
\calL_{\eff}[D_\mu\phi, g_{\mu\nu} ] &= \calL_0[D^\mu\phi D_\mu\phi]
+\calL_2[(D^\nu D_\mu\phi)^2,...]+...
\end{split}
\end{equation}
and the leading order lagrangian ${\cal L}_0$ is an arbitrary  function of the building block  $X= g^{\mu \nu}  D_\mu\phi D_\nu\phi$. In what follows we will focus on the lowest order lagrangian and relate its coefficients to the thermodynamic derivatives.   

\subsection{Thermodynamic matching}

We now relate the functional dependence of ${\cal L}_0 (X)$ on $X$ to the functional 
dependence of the thermodynamic pressure $P(\mu_n)$ on the chemical potential $\mu_n$. 
We refer to this result as to ``thermodynamic matching". 
Although this result is not new~\cite{Son:2002zn,Son:2005rv}, here we provide a derivation that 
can be generalized to other, more complex patterns of symmetry breaking.  In fact,  we will use the generalization of this result in Section~\ref{Section:Mixed} when we consider the simultaneous breaking of translational and particle-number symmetry.

We recall that the thermodynamic interpretation of the functionals  $Z[A^{n}_\mu,g_{\mu \nu}]$ and
$W[A^{n}_\mu,g_{\mu \nu}]$  at constant external fields  ($A_\mu^{n}(x)=\barA_\mu^n$
and $g_{\mu\nu}(x)=\barg_{\mu\nu}=\eta_{\mu\nu}$) is given by the relation
\begin{equation}
\begin{split}
Z_n[\barA_\mu^{n}, \eta_{\mu\nu}]&=e^{iW_n[\barA_\mu^n, \eta_{\mu\nu}]} = e^{-iVT\Omega_n}
=\int [d\phi] \,  e^{i {\cal S}_{\rm eff}  [\barD_\mu\phi, \eta_{\mu\nu}]} ~,\label{eq:GenFuncn}
\end{split}
\end{equation}
where  
$\Omega_n = \langle \Omega | \hat{H} - \barA^n_\mu j_n^\mu |\Omega  \rangle$ is the free energy
density, $V$ is the volume and $T$ is the extent in the time direction.  Here $| \Omega \rangle$ is
the ground state of the Hamiltonian  modified by the presence of the external source $\bar A^n_\mu$
($\hat{H}$ is the Hamiltonian density).  For the specific choice  $\bar A^n_\mu = (m_n + \mu_n,
\vec{0})$ and at zero temperature,   $| \Omega \rangle$ is the many-body ground state at chemical
potential $\mu_n$ and $\Omega_n = \langle \Omega | \hat{H} - (\mu_n+m_n)\hat{j}^0_n |\Omega  \rangle
= - P (\mu_n)$, where $P(\mu_n)$ is the usual thermodynamic pressure of the system.   

The low energy effective action $  {\cal S}_{\rm eff} [\barD_\mu\phi, \eta_{\mu\nu}] $ is a function of the external fields and the Goldstone fields.  
It contains all the quantum dynamics of the high energy Fermionic modes encoded as low energy coefficients. 
In order to evaluate the partition function  $Z[A^{n}_\mu,g_{\mu \nu}]$ we expand the effective action about its saddle point $\phi_0$ which satisfies 
\begin{equation}
\frac{\delta {\cal S}_{\rm eff}[\bar{D}_\mu\phi(x), \eta_{\mu\nu}]}{\delta\phi(x)}|_{\phi_{0}} = 0 
= - \partial_\mu \frac{d \calL_{\eff}(\barD_\mu\phi(x), \eta_{\mu\nu})}{d \partial_\mu\phi(x)}|_{\phi_{0}}\;,\label{eq:phicl}
\end{equation}
minimizes the Euclidean action,  and is well behaved at infinity.  For general external fields
$A_\mu^n(x)$, the solution to Eq.~\ref{eq:phicl} is a functional of the external field,
$\phi_{0}[A_\mu^n]$. However, for our homogeneous and static system with constant external fields the
well behaved solution is $\phi_0=0$.  Expanding about this point we can write 
\begin{equation}
 {\cal S}_{\rm eff} [\barD_\mu\phi, \eta_{\mu\nu}] 
 = {\cal S}_{\rm eff}|_{\phi_{0}=0}  
 + \frac{1}{2}\int d^4 x d^4x' 
 \varphi(x)\varphi(x')
 \frac{\delta^2\cal S_{\rm eff}}{\delta\phi(x)\delta\phi(x')}|_{\phi_{0}}+...\;,
\label{eq:seff_exp}
\end{equation} 
where $\varphi=\phi-\phi_0$, and thus Eq.~\ref{eq:GenFuncn} can be evaluated as a loop expansion 
\begin{eqnarray}
e^{iW[\barA_\mu^n, \eta_{\mu\nu}]}
&=&e^{i {\cal S}_{\rm eff}|_{\phi_{0}=0}+W_{\rm 1-loop}+\cdots}  \\
e^{i W_{\rm 1-loop}} &=&  \int[d\varphi] 
e^{i(\frac{1}{2}\int d^4 x d^4x'\varphi(x)\varphi(x')
\frac{\delta^2\calS_{\rm eff}}{\delta\phi(x)\delta\phi(x')}|_{\phi_{0}}+...)}~, 
\label{eq:WGaussian}
\end{eqnarray} 
where we have explicitly displayed only the quadratic (Gaussian) part of the functional integral in
$[d\varphi]$ which corresponds to
the one-loop approximation. Let us now discuss this loop expansion in light of the EFT power
counting.  The key observation, which is a generic feature of low-energy effective theories~\cite{Weinberg:1978kz}, 
is the following:  within the momentum (gradient) expansion of the EFT, 
loop diagrams generated by $\calL_0$ are higher order than tree-level diagrams with vertices from 
${\cal L}_0$. In our case, one-loop contributions to phonon amplitudes are suppressed by {\it four} powers of momenta
compared to the tree graphs generated by $\calL_0$ as shown by Son and Wingate~\cite{Son:2005rv}.   
Using the above considerations we can write
\begin{equation}
\begin{split}
W[A_\mu^n] &= \int d^4 x\calL_{\eff}\bigl((D_\mu\phi_{0}[A_\mu^n]), \eta_{\mu\nu}\bigr) + W_{\rm{1-loop}}(A_\mu^n) + ...\\
&=\int d^4 x 
\, \Big[ \calL_0\bigl( X_0\bigr) 
+ \calL_2[A_\mu^n] + \calL_4[A_\mu^n]  \Big] 
+ W_{\rm{1-loop}}(A_\mu^n) + ...~,  \label{eq:WloopExpansion}
\end{split}
\end{equation}
where $X_0=D_\mu\phi_0 D^\mu\phi_0$. $\calL_0\bigl( X_0\bigr)$ in Eq.~\ref{eq:WloopExpansion} is the leading term ($\calO(p^0)$),
the second term is of $\calO(p^2)$, the third and fourth are $\calO(p^4)$. The contribution of
$\calO(p^0)$ involves either no derivatives on the external fields or two derivatives compensated by
a Goldstone propagator of $\calO(p^{-2})$. Higher order contributions necessarily involve
derivatives acting on the external field $A_\mu^n(x)$. So we arrive at a very important result: for
very long wavelength external field ($A_\mu^n(x)\rightarrow {\rm{constant}})$, only the first term
in Eq.~\ref{eq:WloopExpansion} survives., i.e.,
\begin{equation}
W[\barA_\mu^n] = \int d^4x \,  \calL_0(\barA_\mu^n \barA^{\mu\;n}) = VT \, 
\calL_0(\barA_\mu^n \barA^{\mu\;n})\label{eq:W4}
\end{equation}
Now recall that  $\barA_\mu \barA^\mu = (m_n + \mu_n)^2$ so that $ \mu_n = \sqrt{\barA_\mu
\barA^\mu} - m_n$.
Moreover, ${\cal L}_0$ depends on $D_\mu \phi$ only through $X$. 
At the classical solution, for  constant external field, one has $X \to X_0  = \barA_\mu \barA^\mu$. 
So we have from Eqs.~\ref{eq:W4} and \ref{eq:GenFuncn}, 
\begin{equation}
{\cal L}_0 (X_0)  =  P (\sqrt{\barA_\mu \barA^\mu} - m_n) =  P (\sqrt{X_0} - m_n = \mu_n) ~.
\label{eq:W6}
\end{equation}
The above  relation fixes the functional dependence of ${\cal L}_0$ on the variable $X$ 
once the functional dependence of $P$ on $\mu_n$ is known. 
So  in general we have:
\begin{equation}
{\cal L}_0 (X) = P(Y \equiv \sqrt{X} - m_n)  ~.
\label{eq:LisP} 
\end{equation}
Finally we note that in the non-relativistic limit the relevant building block takes the form 
$Y=  \sqrt{(m_n+\mu_n+\partial_0\phi)^2-(\partial_i\phi)^2}-m_n\
\simeq\mu_n+\partial_0\phi-\frac{(\partial_i\phi)^2}{2m_n}$~\cite{Son:2002zn,Son:2005rv,Greiter:1989qb}.

\subsection{Identifying the low-energy constants}

Eq.~\ref{eq:LisP} gives us the complete expression of the superfluid lagrangian to the lowest order in $(m-n)$.
Expanding the function $\calL_0(\phi)$ in powers of the Goldstone fields, one can read off 
the phonon kinetic term and self-interaction vertices:
\begin{equation}
\begin{split}
{\cal{L}}_{\eff}[\phi] 
&= \calL_0[(\mu_n+m_n)^2]+  \frac{d P}{d Y}  \,  (\partial_0\phi)
+\frac{1}{2!} \frac{d^2 P}{d Y^2}  \, (\partial_0\phi)^2
-\frac{1}{2!(\mu_n+m_n)} \frac{\partial P}{\partial{Y}} \,  (\partial_i\phi)(\partial_i\phi)\\
&+\frac{1}{3!}  \frac{d^3 P}{d Y^3} \, (\partial_0\phi)^3
+\frac{1}{2!} \left[ -\frac{1}{(\mu_n+m_n)}\frac{d^2 P}{d Y^2}
+\frac{1}{(\mu_n+m_n)^2}\frac{d P}{d Y}\right](\partial_0\phi)(\partial_i\phi)(\partial_i\phi)
+...~\label{eq:LowEnergyAction}
\end{split} 
\end{equation}
where 
all derivatives are evaluated at $Y=\mu_n$.  The above expansion makes it clear that the low energy
constants of the theory are then given by the derivatives of the pressure with respect to the
chemical potential. Eq.~\ref{eq:LowEnergyAction} also serves as a starting point for a fluid
dynamical study of the system once one realizes that $-\partial_i\phi/(m_n+\mu_n)$ is the velocity
of the superfluid. (See Ref.~\cite{Mannarelli:2009ia} and references therein.)

Alternatively,  one can show that the thermodynamic derivatives are related to the
static correlation functions involving the neutron charge and current operators. 
For example,
\begin{equation}
\frac{\partial W_n}{\partial \barA_{\mu}^{n}}
=\langle J_n^\mu(0)\rangle 
=\frac{d P}{d Y}|_{eq}\eta^{\mu 0}\;.
\end{equation}
Similarly,
\begin{equation}
\begin{split}
\frac{\partial^2 W_n}{\partial \barA_{\nu}^{n}\partial \barA_{\mu}^{n}}|_{eq}
&= i\int d^4 x \langle T\{J_n^\nu(x)J_n^\mu(0)\}\rangle_{\rm{connected}}\\
&=\left[ \frac{1}{(\mu_n+m_n)}\frac{d P}{d Y}|_{eq}\eta^{\mu \nu}
+\left(-\frac{1}{\mu_n+m_n}\frac{d P}{d Y}+\frac{d^2 P}{d Y^2} \right)|_{eq}\eta^{\mu 0}\eta^{\nu 0}
\right] \label{eq:F_AA}\;.
\end{split}
\end{equation}
A more careful analysis taking $\barA$ to be long wavelength but not quite constant shows that the
current-current correlation function thus obtained is the momentum independent part of the
transverse-current---transverse-current correlation function.

Finally, in the non-relativistic limit, separating the space and time components we obtain the well known
results
\begin{equation}
\begin{split}
\frac{\partial^2 W_n}{\partial \barA_{0}^{n}\partial \barA_{0}^{n}}|_{eq} &= 
\frac{d^2 P}{d Y^2}\\
\frac{\partial^2 W_n}{\partial \barA_{i}^{n}\partial \barA_{j}^{n}}|_{eq} 
&= \frac{1}{m_n}\frac{d P}{d Y}\eta^{ij}\;.
\end{split}
\end{equation}

\section{The superfluid and lattice phonon lagrangian~\label{Section:Mixed}}

\subsection{Fields and the effective lagrangian} 

A crystalline ground state of the system spontaneously breaks translations and
rotations.  The proton number density acts as the relevant order parameter. Since
space-time dependent translations include rotations as special cases, it will suffice to
consider the breaking of the abelian group $G$ of spatial translations to the subgroup $H = \{
T_{\vec{b}}\}$ containing discrete translations by multiples of lattice basis vectors, ${\vec{b}}$.
The  generators of $G$ are the components of the  three-momentum  $P^a$
given by space integrals of the energy-momentum tensor components
$T^{0a} (x)$, and momentum conservation takes the local form 
\begin{equation}
\label{eq:cc1}
\partial_\mu  \, T^{\mu a} (x) = 0 .
\end{equation}

The Goldstone effective fields can be chosen as space-time dependent  
coordinates $\xi_a (x)$  ($a=1,2,3$) of the coset space $G/H$:  
\begin{equation}
\gamma(x) =  e^{i \xi^a (x)  \, P^a} ~,  \qquad  \gamma \in G/H ~. 
\label{eq:lphdef}
\end{equation}
The {\it nonlinear}  action of the translations group on the Goldstone fields is specified by: 
\begin{equation} 
x_b  \to   x_b'  = x_b + a_b,\;\;
\xi_b (x)    \to     \xi_b ' (x')   =   \xi_b  (x)  + a_b    ~, 
\label{eq:global}
\end{equation}
as can be verified   by left multiplication of  $\gamma \in G/H$ with 
$g = e^{i  a^b P^b} \in G$.

Promoting the global symmetry to local symmetry, a generally covariant
formulation~\cite{Leutwyler:1996er, Son:2002zn} of the phonon dynamics in the
background metric $g_{\mu \nu}$ can be readily achieved by introducing a set of fields that
transform as scalar fields under the general coordinate transformations
of Eqs.~\ref{eq:gr1}:
\begin{equation}
z^{a} (x) = x^a  - \xi^a (x)  \qquad \qquad  a=1,2,3\label{eq:z_a}\;. 
\end{equation}
The fields $z^a (x^\mu)$ can be thought of as 
one particular choice of  body-fixed coordinates \footnote{These are the
coordinates in a frame frozen in the body of the solid. If one follows a material point in the
solid, its coordinates in this frame remain constant.}  of a material point located at $x^\mu =
(t,\vec{x})$. With this choice the body-fixed coordinates coincide with the 
``laboratory" coordinates $(x^a, g_{ab})$ when the displacement field 
$\xi^a$ vanishes.   

As in the pure neutron case,  the partition function in 
the presence of generic external fields $A_\mu^{n}(x)$, $A_\mu^{p}(x)$, $g_{\mu\nu}(x)$,
admits a low-energy representation in terms of the four Goldstone modes 
$\phi$ and $\xi^a$: 
\begin{equation}
Z[A_\mu^{n},A_\mu^{p},g_{\mu\nu}]=
\int  [d \Psi_n][d \Psi_p]  e^{i {\cal {S}}[\Psi_n,\Psi_I,A_\mu^{n},A_\mu^{p},g_{\mu\nu}]}
\ \rightarrow  \ 
\int  [d \phi][d\xi^a]  e^{i {\cal {S}_{\rm eff}}[\phi,\xi^a,A_\mu^{n},A_\mu^{p},g_{\mu\nu}]}~.
\label{eq:Zmixed} 
\end{equation}
At the end, we will evaluate the partition function for space-time independent external fields
$\barA_\mu^n$, $\barA_\mu^p$ and $\barg_{\mu\nu}$, specifying a particular density, and lattice
shape for the system.

${\cal{S}_{\rm eff}}$ represents the effective action of $\xi^a$ (or equivalently $z^a$) and $\phi$ 
in the presence of external fields. We can organize the terms in ${\cal S}_{\rm eff}$ 
according to the same power counting introduced earlier in our discussion of the superfluid, i.e. 
in increasing difference between the number of derivative operators and  the Goldstone fields, 
\begin{eqnarray}
{\cal S}_{\rm eff}  [\phi,\xi^a, A_\mu^{n},A_\mu^{p},g_{\mu\nu}] 
&=& \int d^4 x\sqrt{-g} \Big[  \calL_{0} (\partial_\mu\phi,\partial_\mu z^a,A_\mu^{n},A_\mu^{p},g_{\mu\nu})
\\
&+&
\calL_1(D_\nu \partial_\mu\phi,D_\nu \partial_\mu{z}^{a}, D_\mu A^n_\nu ...)+ ... \Big] ~. 
\end{eqnarray}

Symmetries impose powerful constraints on the form of $\calL_{\eff}$. Since $z^a$ transform as scalars,
$\partial_\mu z^a$ transforms as a contravariant vector. The building blocks of the scalar
function $\calL_{\eff}$ are scalar combinations of ${A}_{\mu}^n(x)$, ${A}_{\mu}^p(x)$, ${g}_{\mu\nu}(x)$, 
$\partial_\mu\phi$, $\partial_\mu {z}^{a}$, and their covariant derivatives. Symmetry under phase
rotations of the neutrons, Eq.~\ref{eq:gauge_n}, implies that $A_{\mu}^n(x)$ should appear in a
combination such that the transformation ${A}_{\mu}^n(x) \rightarrow {A}_{\mu}^n(x)
+\partial_\mu\theta_n(x)$ leaves the effective action invariant. The same is required for the
protons. In the pure neutron case we found that gauge symmetries implied that 
$\partial_\mu\phi$ and ${A}_{\mu}^n(x)$ could appear only in the combination $D_\mu\phi$. 
To lowest order in the power counting, 
the scalar combinations that can be constructed  from the gauge invariant combinations are
\begin{eqnarray}
X &=& g^{\mu \nu} D_\mu\phi D_\nu\phi \\
W^a&=& g^{\mu \nu}  \, D_\mu\phi \partial_\nu  z^a   \\
H^{ab}&=& g^{\mu \nu}  \, \partial_\mu z^a \partial_\nu z^b~.
\end{eqnarray}
In addition to these building blocks, other possibilities arise in the mixed case 
that were not present in the case of a pure neutron superfluid. 
The following terms only change by a total derivative on making gauge transformations defined in Eqs.~\ref{eq:gauge_n}, \ref{eq:gauge_I},  
\footnote{The term
\begin{equation}
\int d^4x \frac{C_3}{3!} \epsilon^{\mu\nu\sigma\lambda}
\epsilon^{abc}\Bigl(A^{n}_{\mu}(x) \Bigr)(\partial_\nu z^a(x)\partial_\sigma z^b(x)\partial_\lambda z^c(x))
\end{equation}
can be rewritten as,
\begin{equation}
\int d^4x \frac{C_3}{3!}\Bigl[ 3! \sqrt{g\det\bigl(\IH\bigr)}-\epsilon^{\mu\nu\sigma\lambda}
\epsilon^{abc}
\Bigl(\partial_{\mu}\phi(x) \Bigr)(\partial_\nu z^a(x)\partial_\sigma z^b(x)\partial_\lambda z^c(x))
\Bigr]\;,
\end{equation}
where,
\begin{equation}
\IH = \left[\begin{array}{cc}
X & W^{aT}\\
W^{a} & H^{ab}
\end{array}
\right]~.
\end{equation}
This shows that any term proportional to $C_{3}$ can be reabsorbed by  a redefinition of the function $f$ and the coefficient $C_{2}$.} 
\begin{equation}
\int d^4x \frac{1}{3!} \epsilon^{\mu\nu\sigma\lambda}
\epsilon^{abc}\Bigl(C_1 \,  A^{p}_{\mu}(x) + C_2   \, \partial_\mu\phi(x) \Bigr)(\partial_\nu z^a(x)\partial_\sigma z^b(x)\partial_\lambda z^c(x))\;. 
\end{equation}
Hence, the most general form of $\calL_0$ is 
\begin{equation}
\begin{split}
\calL_0(\partial_\mu\phi, &\partial_\mu z^a, A^n_\mu, A^p_\mu,g_{\mu\nu} ) 
= 
 f(X, W^a, H^{ab} ) \\
&+ \frac{1}{3!\sqrt{-g}}
\epsilon^{\mu\nu\sigma\lambda}\epsilon^{abc}
\Bigl(C_1  \,  A^p_{\mu} + C_2  \, \partial_\mu\phi\Bigr)(\partial_\nu z^a\partial_\sigma z^b\partial_\lambda z^c)
~. 
\end{split}
\end{equation}
The term proportional to $C_2$ is a total derivative 
that becomes relevant only in presence of non-trivial topological configurations 
(vortices)  for the field $\phi$~\cite{Son:2005ak}. This would be relevant for calculations of vortex-phonon interactions
but for now on we disregard this term and restrict our discussion to vortex free configurations.

\subsection{Thermodynamic matching}
Extending the analogy with the neutron superfluid case further we can relate ${\calL}_0$ to the
free energy of the neutron-proton system. 
The free energy $\Omega[A_\mu^n, A_\mu^p, g_{\mu\nu}]$ is proportional to the $\log$ of the 
partition function. Following the discussion  in the pure neutron case,  one can show that 
for  space-time independent  external fields, $g_{\mu\nu}(x)=\barg_{\mu\nu}$, $A_\mu^{p}(x)=\barA_\mu^{p}$,
$A_\mu^{n}(x)=\barA_\mu^n$, 
$\Omega$ 
it is also equal to $\calL_0$  evaluated at the classical solution
$\phi|_{0}=0$, $\xi^a|_{0}=0$. Hence,
\begin{equation}
\begin{split}
Z[\barA_\mu^n, \barA_\mu^p, \barg_{\mu\nu}]&=e^{iW[\barA_\mu^n, \barA_\mu^p, \barg_{\mu\nu}]}=
e^{-iVT\Omega[\mu_n, \mu_p, \barg_{\mu\nu}]}  
=e^{iVT\calL_0(0,\delta_\mu^a, \barA_\mu^n, \barA_\mu^p, \barg_{\mu\nu})}\;,
\label{eq:barOmega}
\end{split}
\end{equation}
where $VT=\int d^4x\sqrt{-\barg}$. 
For this choice of the many body ground state, $X_0=\barA_\mu^n\barA^{\mu\;n}$, $W^a_0=0$ and $H^{ab}=\barg^{ab}$. Therefore,
\begin{equation}
\begin{split}
-\Omega[\mu_{n},\mu_{p}, \barg_{\mu\nu}]
&=f(X=X_0, W^a=0, H^{ab}=\barg^{ab})+\frac{1}{\sqrt{-\barg}}C_1 \, (\mu_p+m_p)\;.
\end{split}
\end{equation}

The constant $C_1$ can be determined from the requirement that
$\frac{\partial\Omega}{\partial\mu_p}=\frac{C_1}{\sqrt{-\barg}}_p$. Thus, we see that $C_1$ is the
density of protons for a configuration whose metric has determinant $-1$. Symbolically,
$C_1=n_p^\eta$, where $\eta_{\mu\nu}$ is a particularly convenient choice (also see footnote
$5$) for a metric with determinant $-1$.

When we consider the functional form of $f$, we encounter a feature different from the previous
case where we considered the pure neutron superfluid. There, we were able to
determine the complete dependence of the function $f$ on its arguments from the free energy
function $\Omega$ ($\calL_0(X)=P(Y)=-\Omega_n(Y)$), i.e. from
a calculation of the partition function with the specific form 
$\barA_\mu^n = (m_n + \mu_n, \vec{0})$ 
for the external field. 
In the mixed case, however, since $D_\mu\phi \partial^\mu
z^a|_{eq} = 0$ for $\barA_\mu^n = (m_n + \mu_n, \vec{0})$,   
it is not possible to calculate the dependence of $f$ on $W^a$  
from the free energy calculation in this external field. \footnote{This fact is intuitively 
understandable. In the non-relativistic limit~\cite{Son:2002zn} we have
$W^{a}\sim m_n(-\frac{1}{m_n}\partial_a\phi-\partial_0\xi^a+\frac{1}{m_n}\partial_i\phi\partial_i\xi^a)
= m_n(v_n^a-\partial_0\xi^a-v_n.\nabla\xi^a)$ which is the relative velocity between the neutron
superfluid and the proton clusters. The dependence on $W^a$ therefore represents the interaction
between the superfluid neutrons and the lattice when they are moving relative to each other, and can
not be calculated by a ground state evaluation of the free energy.}  
To determine  the dependence of $f$ on $W^a$  one needs to evaluate the partition function $Z$ for a
space-time independent external gauge field $A_\mu^n(x)$ that has non-zero spatial components,
$\tilA_\mu^{n}=(\mu_n+m_n,{\bf{A}}_i)$. This gives $\tilD_\mu\phi_{0}= \tilA_\mu^{n}$,
$\tilde{X}_{0}=\tilA_\mu^n\tilA^{\mu\;n}$ and $\tilde{W}^a_0=\barg^{a\nu}\tilA_\nu^n=\tilA^{a\;n}$. 
Then, 
\begin{eqnarray}
-\Omega [\tilA_\mu^{n},\barA_\mu^{p}, \barg_{\mu\nu}]
&=&  \calL_0 (0,\delta^a_\mu,\tilA_\mu^n,\barA_\mu^{p}, \barg_{\mu\nu}) \\
&=& f(X=\tilde{X}_{0}, W^a=\tilde{W}^a, H^{ab}=\barg^{ab})+\frac{1}{\sqrt{-\barg}}n^{\eta}_p(\mu_p+m_p)   
\end{eqnarray}
By calculating the free energy for various $\tilA_\mu^n$ and $\barg^{ab}$ we can map out the
functional dependence of $f$ on $X$, $W^a$ and $H^{ab}$. Finally, noting that
$\Omega=\langle\Omega| \hat{H} - \tilA_\mu^n j^\mu_n - n_p(\mu_p+m_p)|\Omega\rangle$ ($\hat{H}$ is 
the hamiltonian density),  the term
proportional to $n_p$ cancels out from both sides and the function $f$ is given by,
\begin{equation}
\begin{split}
f(\tilde{X}_{0}, \tilde{W}^a_0, \barg^{ab})=\langle\Omega|\tilA_\mu^n j^\mu_n-\hat{H} |\Omega\rangle\;.
\end{split}
\end{equation}
The generalization of Eq.~\ref{eq:barOmega} with the full $\tilA_n$ is simply 
\begin{equation}
\begin{split}
Z[\tilA_\mu^n, \barA_\mu^p, \barg_{\mu\nu}]&=e^{iW[\tilA_\mu^n, \barA_\mu^p, \barg_{\mu\nu}]}=
e^{-iVT\, \Omega[\tilA_\mu^n, \mu_p, \barg_{\mu\nu}]}  
=e^{iVT\calL_0(0, \delta_\mu^a, \tilA_\mu^n, \barA_\mu^p, \barg_{\mu\nu})}\;.
\label{eq:tilOmega}
\end{split}
\end{equation}

\subsection{Identifying the low-energy constants}

Expanding the function $\calL_0$ in powers of the Goldstone fields $\phi$ and $\xi^a$,  
one can read off  the phonon kinetic term (including kinetic mixing 
among the $\xi$ and $\phi$) and self-interaction vertices.  
The expansion in the $\phi$ field can be done as in Section~\ref{Section:Neutron}. 
The expansion in $\xi^a$ is performed about the undeformed equilibrium 
configuration with $\barg_{\mu\nu}=\eta_{\mu\nu}$ and $\xi^a=0$.  
Deviations from the equilibrium shape are then signified by
$\xi^a\neq 0$, which gives $H^{ab}=\eta^{ab}+\Delta {H}^{ab}$ with
$\Delta {H}^{ab}=-(\partial^a\xi^b+\partial^b\xi^a)+\partial_\mu\xi^a\partial_\nu\xi^b\eta^{\mu\nu}$. At
equilibrium, $X=X_0$ and $W^a=W^a_0=0$, and deviations from equilibrium are given by $X-X_0=\Delta
X=2(\mu_n+m_n)\partial_0\phi+\partial_\mu\phi\partial_\nu\phi\eta^{\mu\nu}$ and $W^a-W^a_0=\Delta
W^a=-(\mu_n+m_n)\partial_0\xi^a+\eta^{ab}\partial_b\phi-\partial_\mu\phi\partial_\nu\xi^a\eta^{\mu\nu}$. 

To second order in the fields, the expansion of $f$ is
 \begin{equation}
 \begin{split}
 f(X, W^a, H^{ab})
 &= f(X_0, {\bf{0}}, \eta^{ab})\\
 &+\frac{\partial f}{\partial X}\Big|_{eq}
  (2(\mu_n+m_n)\partial^0\phi+\partial_\mu\phi\partial^\mu\phi) 
  +\frac{1}{2!}\frac{\partial^2 f}{\partial X^2}\Big|_{eq}
  (2(\mu_n+m_n)\partial^0\phi)^2 \\
 &+\frac{1}{2}\frac{\partial^2 f}{\partial W^a\partial W^b}\Big|_{eq}
  ({-}(\mu_n+m_n)\partial^0\xi^a+\partial^a\phi)({-}(\mu_n+m_n)\partial^0\xi^b+\partial^b\phi) \\
 &+\frac{\partial f}{\partial H^{ab}}\Big|_{eq}
   \Delta{H}^{ab}+\frac{1}{2}\frac{\partial^2 f}{\partial H^{ab} \partial H^{cd}}
   \Delta{H}^{ab}\Delta{H}^{cd}\\
 &+\frac{\partial^2 f}{\partial H^{ab} \partial X}\Big|_{eq}
   \Delta{H}^{ab}(2(\mu_n+m_n)\partial^0\phi)~. 
 ~\label{eq:lowElagrangian2}
 \end{split}
 \end{equation}
We have simplified the expansion above by taking $\frac{\partial f}{\partial W^a}|_{eq}=0$,
$\frac{\partial^2 f}{\partial X\partial W^a}|_{eq}=0$ and $\frac{\partial^2 f}{\partial H^{ab}
\partial W^c}|_{eq}=0$. This would be the case for any crystal with reflection symmetry, for
example a cubic crystal.
For a cubic crystal, one can further simplify the expressions by using symmetry under rotation by
$\frac{\pi}{2}$ along the axes. This gives $\frac{\partial^2 f}{\partial W^a\partial W^b} =
\frac{1}{3}\frac{\partial^2 f}{\partial W^c\partial W^c}\delta^{ab}$, $\frac{\partial  f}{\partial
H^{ab}}=\frac{1}{3}\frac{\partial  f}{\partial H^{cc}}\delta^{ab}$, and $\frac{\partial^2  f}{\partial X\partial
H^{ab}}=\frac{1}{3}\frac{\partial^2 f}{\partial H^{cc}\partial X}\delta^{ab}$, where
$\frac{\partial}{\partial H^{cc}}=\frac{\partial}{\partial H^{11}}\!+\frac{\partial}{\partial
H^{22}}+\!\frac{\partial}{\partial H^{33}}\!$. Finally, we make 
the non-relativistic approximation {$\mu_n\ll m_n$, $\mu_p\ll m_p$,} and separate the space and time
components.

With all these simplifications,
\begin{equation}
\begin{split}
f(&X, W^a, H^{ab})
= f(X_0, {\bf{0}}, \eta^{ab}) + \frac{1}{2}\frac{\partial^2 f}{\partial Y^2}(\partial_0\varphi)^2
   +\frac{1}{2}\partial_i\varphi\partial_j\varphi\eta^{ij}
   \left[  \frac{1}{m_n}\frac{\partial f}{\partial Y}-\frac{\partial^2 f}{3\partial W^c\partial W^c} \right] \\
&+\frac{1}{2} \left[ \frac{2}{3}\frac{\partial f}{\partial H^{cc}}+m_n^2\frac{\partial^2 f}{3\partial W^c\partial W^c} \right]\partial_0\xi^a\partial_0\xi^a
   +(\partial^0\varphi \partial_a\xi^a)  \left[ \frac{2}{3}\frac{\partial^2 f}{\partial H^{ee} \partial Y} +
   m_n\frac{\partial^2 f}{3\partial W^c\partial W^c} \right]\\
&+\frac{1}{3}\frac{\partial f}{\partial H^{cc}} (\partial_i\xi^a\partial^i\xi^a)
   +\frac{1}{2}\frac{\partial^2 f}{\partial H^{ab} \partial H^{cd}}
   (\partial_a\xi^b+\partial_b\xi^a)(\partial_c\xi^d+\partial_d\xi^c)+...
 ~\label{eq:lowElagrangianinterm}
 \end{split}
 \end{equation}

 One key consequence of  Eq.~\ref{eq:lowElagrangianinterm}
is that the low-energy constants are related to derivatives of the 
function $f$ with respect to  $X$, $W^a$, and $H^{ab}$ evaluated at the 
``equilibrium"  point  $X=X_0,\;W^a=W^a_0,\;H^{ab} = \eta^{ab}$.
In turn, due to the thermodynamic matching relation Eq.~\ref{eq:tilOmega}, 
the low-energy constants can be expressed in terms of 
derivatives of the generating functional $W[\tilA_\mu^n, \barA_\mu^p, \barg_{\mu\nu}]$.  
The analysis proceeds along parallel lines to the pure neutron case,  
but it  contains a number of novel features, which we discuss in some detail below.

\subsubsection{Thermodynamic derivatives}

The first order derivatives  of the functional $W[\tilA_\mu^n, \barA_\mu^p, \barg_{\mu\nu}]$
specify the number density of
particles in, and the stress energy tensor of, the system:
\begin{equation}
\begin{split} 
\frac{1}{VT}\frac{\partial W}{\partial A_0^{n}}|_{eq}
  =\langle n_n\rangle
  &= \frac{\partial f}{\partial Y} 
  \\
\frac{-2}{VT}\frac{\partial W}{\partial g_{00}}|_{eq}
  =\langle T^{00}\rangle 
  &=(m_n + \mu_n) \, \frac{\partial f}{\partial Y}-f
  \\
\frac{-2}{VT}\frac{\partial W}{\partial g_{ab}}|_{eq}
 = \langle T^{ab}\rangle
 &=\bigl[-2\frac{1}{3}\frac{\partial f}{\partial H^{cc}}-f\bigr]\eta^{ab}
~\label{eq:W1derivatives}\;.
\end{split}
\end{equation}
In particular, $2\frac{1}{3}\frac{\partial f}{\partial H^{cc}} = -\frac{1}{3}\langle T^a_a\rangle -
f$. The second order derivatives of $W$ have the following form, 
\begin{equation}
 \begin{split} 
\frac{1}{VT}\frac{\partial^2 W}{\partial A_0^{n}\partial A_0^{n}}|_{eq}&=
  \frac{\partial^2 f}{\partial Y^2}   \equiv -F_{A_{0}A_{0}}\\
\frac{1}{VT}\frac{\partial^2 W}{\partial A_a^{n}\partial A_b^{n}}|_{eq}&=
  \bigl[\frac{1}{m_n}\frac{\partial f}{\partial Y}
  -\frac{1}{3}\frac{\partial^2 f}{\partial W^c\partial W^c}\bigr]\eta^{ab} \equiv -F_{A_{a}A_{b}}\\
\frac{1}{VT}\frac{\partial^2 W}{\partial A_0^{n}\partial g_{00}}|_{eq}&=m_n\frac{\partial^2
f}{\partial Y^2} \equiv -F_{A_{0}g_{00}}\\
\frac{-2}{VT} \frac{\partial^2 W}{\partial A^n_0\partial g_{ab}}|_{eq}&=
  \bigl[{-\frac{\partial f}{\partial Y}
  -2\frac{1}{3}\frac{\partial^2 f}{\partial H^{cc}\partial Y}}\bigr]\eta^{ab}  \equiv -F_{A_{0}g_{ab}}\\
\frac{-2}{VT} \frac{\partial^2 W}{\partial A_a^{n}\partial g_{b0}}|_{eq}&=
  \bigl[\frac{\partial f}{\partial Y}-{{m_n}}
  \frac{\partial^2 f}{3\partial W^c\partial W^c}\bigr]\eta^{ab}  \equiv -F_{A_{a}g_{b0}}\\
\frac{4}{VT} \frac{\partial^2 W}{\partial g_{00}\partial g_{00}}|_{eq}&=
  {-f+m_n \frac{\partial f}{\partial Y}+m_n^2\frac{d^2 f}{d Y^2}} \equiv -F_{g_{00}g_{00}}\\
\frac{4}{VT}\frac{\partial^2 W}{\partial g_{00}\partial g_{ab}}|_{eq}&=
  \bigl[-f - m_n\frac{\partial f}{\partial Y}
  + 2\frac{1}{3}\frac{\partial f}{\partial H^{cc}}
  - 2\frac{1}{3} {m_n}\frac{\partial^2 f}{\partial Y\partial H^{cc}}\bigr]\eta^{ab} \equiv -F_{g_{00}g_{ab}}\\
\frac{4}{VT}\frac{\partial^2 W}{\partial g_{a0}\partial g_{b0}}|_{eq}&=
  \bigl[-f + m_n\frac{\partial f}{\partial Y}
  - 2\frac{1}{3}\frac{\partial f}{\partial H^{cc}}
  - {m_n^2}\frac{\partial^2 f}{3\partial W^c\partial W^c}\bigr]\eta^{ab} \equiv -F_{g_{a0}g_{b0}}\\
\frac{4}{VT}\frac{\partial^2 W}{\partial g_{ab}\partial g_{cd}}|_{eq}&=
  {-(f+ \frac{4}{3}\frac{\partial f}{\partial
  H^{ee}})(\eta^{ac}\eta^{bd}+\eta^{ad}\eta^{bc}-\eta^{ab}\eta^{cd})
  +4\frac{\partial^2 f}{\partial H^{ab}\partial H^{cd}}}
  \\
  & 
  \equiv -F_{g_{ab}g_{cd}}
\;,
~\label{eq:W2derivatives}
 \end{split}
 \end{equation}
where all derivatives of $f$ are evaluated at equilibrium. The second order correlations are
proportional to the momentum independent part of appropriate time ordered correlation functions of the
neutron current and the total stress energy tensor. Since they can be found simply from 
Eq.~\ref{eq:W2derivatives} by noting that $j^\mu$ and $T^{\mu\nu}$ are  obtained as the partial derivatives of $W$ with respect to $A_{\mu}$ and $g_{\mu\nu}$, respectively, we don't explicitly include them here. The expressions in 
Eqs.~\ref{eq:lowElagrangian2} and \ref{eq:W2derivatives} look complicated  but have simple physical 
interpretations, as we discuss below.

\subsubsection{The entrainment coefficient} 
\label{sec:entrainment}
Here we note that the current-current correlation function for neutrons, $\frac{\partial^2
W}{\partial A_a^{n}\partial A_b^{n}}|_{eq}$ is not simply proportional to the total neutron density
but is instead proportional to $\frac{d f}{d Y}-\frac{m_n}{3} \frac{\partial^2 f}{\partial
W^c\partial W^c}$ .  We conjecture that $n_b \equiv \frac{m_n}{3}\frac{\partial^2 f}{\partial
W^c\partial W^c}>0$ and this represents the number density of neutrons ``bound'' or ``entrained'' on
the nuclei. $\frac{\partial f}{\partial Y}-m_n\frac{1}{3}\frac{\partial^2 f}{\partial W^c\partial
W^c}$ is then interpreted as the number of ``unbound'' neutrons in the system.  Indeed, from the
coefficient of $(\partial_i \phi)^2$ in Eq.~\ref{eq:lowElagrangianinterm}  we see that the current
of the superfluid mode is proportional to $n_f=(n_n-n_b)$, where $n_f$ is neutron density that can
participate in superfluid transport.

It is also reassuring that in Eq.~\ref{eq:lowElagrangianinterm} the effective mass density of the
``proton'' clusters involved in lattice vibrations is correspondingly increased by $m_n n_b$.  This is easily seen by noting that the coefficient of the kinetic term is given by $\frac{1}{2}[\frac{2}{3}\frac{\partial f}{\partial H^{cc}}
+m_n^2\frac{1}{3}\frac{\partial^2 f}{\partial W^c\partial W^c}]
=\frac{1}{2}[\frac{2}{3}\frac{\partial f}{\partial H^{cc}}+m_nn_b]$.  The picture of the inner crust as periodic
clusters of ions and neutrons ``entrained'' on the clusters has been discussed
previously~\cite{Carter:2004pp,Carter:2006}. Our formalism confirms this intuition, and provides a field theoretic derivation of the 
entrained neutron density in terms of generalized thermodynamic derivatives.

\subsubsection{Relating the LECs to the stress and elastic tensors}

{From Eq.~\ref{eq:lowElagrangian2} and the last relation in Eq.~\ref{eq:W2derivatives} 
one sees that the part of the effective lagrangian  quadratic in gradients of $\xi^a$ 
can be expressed in terms of $\frac{\partial^2 W}{\partial g_{ab}\partial g_{cd}}|_{eq}$.
This result establishes a non-trivial relation between the LECs  
appearing in the phonon quadratic lagrangian and the 
stress tensor correlator that can be calculated in the underlying  theory  using non-perturbative  methods.  
We can go one step further and relate the LECs to first and second order thermodynamic derivatives  
of the free energy with respect to the strain tensor (i.e. the stress and elastic tensors).  
This step relies on the relationship between the external metric  
$g_{\mu\nu}$ and the  strain of the crystal structure, i.e. its shape. 
We will find that since the strain has pieces both linear and quadratic in the
displacement fields (see Eq.~\ref{eq:strain} below) the elastic constants are linear combinations of first
and second order derivatives of the free energy with respect to the strain.}


Let us first  recall a few basic definitions from the theory of elasticity.  The elastic constants
can be defined through thermodynamic derivatives of the free-energy (or internal energy)  per unit
mass \footnote{Equivalently, one defines the thermodynamic quantities per unit volume of the
{\it{undeformed}}~\cite{Landau:Mechanics} system.  The free energy per unit flat space volume
element is given by $F=\sqrt{-\barg} \Omega=-\frac{\sqrt{-\barg}}{VT}W[\tilA_\mu^n, \barA_\mu^p,
\barg_{\mu\nu}]$.  } with respect to the strain tensor $s_{ab}$  associated with deformations around
some reference point: 
\begin{equation}
F = F_0-t^{ab} s_{ab}+ \frac{1}{2}\, X^{abcd} \  s_{ab} s_{cd}    \ + \  ... 
\label{eq:Fexp}
\end{equation}
In the above relation $t^{ab}$ is the stress tensor associated with the reference configuration,
which we will take to be the equilibrium configuration. 
{$X^{abcd}$ is known as the elastic tensor.}
For an equilibrium configuration in absence 
of external forces, $t^{ab}=0$,  and {the components of}  $X^{abcd}$ are simply the elastic constants. 
For an equilibrium
configuration in the presence of external forces ($t^{ab}\neq0$), {for example a solid under pressure}, 
the elastic constants are linear combinations of $t^{ab}$ and $X^{abcd}$. 

The strain tensor is defined in terms of the displacement fields $\zeta^a(x)$ as follows:
\begin{equation}
s_{ab} = \frac{1}{2} \left( \frac{\partial \zeta^a}{\partial x^b}  + \frac{\partial \zeta^b}{\partial x^a} 
 +  \frac{\partial \zeta^i}{\partial x^a} \frac{\partial \zeta^i}{\partial x^b}  \right)~. 
\label{eq:strain}
\end{equation}
The strain tensor has a simple geometric interpretation. 
Imagine  choosing the body-fixed coordinates $x^a$  so that they coincide with the Euclidean (flat)  
laboratory coordinates  when the body is in equilibrium.  
After a deformation specified by the displacement fields $\zeta^a(x)$, 
the body-fixed coordinate system will have a non-trivial three-dimensional 
metric, whose deviation from flat metric is specified by $s_{ab}$:
\begin{equation}
\eta_{ab} \equiv -\delta_{ab}   \to    g_{ab} = \eta_{ab}  - 2 \, s_{ab}~.
\label{eq:strain2}
\end{equation}

We now state the results that ensure the connection with the elastic constants, 
relegating their proof to Appendix~\ref{sec:app1}.   
The main point is that the  energy density   $\Omega [\tilA^n,\barA^p,\barg]$  calculated using the path
integral (Eqs.~\ref{eq:Zmixed} and \ref{eq:barOmega}) 
 in the presence of a space-time independent metric of block form
\begin{equation}
\barg_{\mu\nu} = 
\left[\begin{array}{cc}
1 & 0\\
0 & \barg_{ab}
\end{array}\right]~\label{eq:g0} 
\end{equation}
is equal to the flat-space 
energy density 
in the lowest energy state  $| \Omega_g \rangle $  subject to the 
``deformation condition" 
$\langle \Omega_g | \hat{\xi}^a (x) | \Omega_g \rangle =  \zeta_g^a (\vec{x})$, 
with $\zeta^a_g(x)$ 
related to $\barg_{ab}$ by
\begin{equation}
\barg_{ab}=\eta_{ab} - 2 \,  s_{ab} (\zeta_g) 
~\label{eq:gabvsxi}\; ,
\end{equation}
and $s_{ab} (\zeta)$ given in Eq.~\ref{eq:strain}.
This result establishes a correspondence between 
the ground state in presence of $\barg_{ab} \neq \eta_{ab}$  
and a  deformed configuration around the ``true ground state" in 
the absence of external gravitational field, $\barg_{ab}=\eta_{ab}$.  Therefore, by varying the external
metric $\barg_{ab}$ we probe different deformed configurations of the system,  with strain tensor   
related to $\barg_{ab}$  by Eq.~\ref{eq:gabvsxi}.

Identifying the free energy $F[s]$ per unit volume in flat space  with 
$F=\sqrt{-\barg} \Omega=-\frac{\sqrt{-\barg}}{VT}W[\tilA_\mu^n, \barA_\mu^p, \barg_{\mu\nu}]$ 
we have:
\begin{equation}
X^{abcd} \equiv  \frac{\delta^2  F}{\delta s_{ab} \delta s_{cd}} \Bigg|_{s=0} = 
 \frac{\delta^2  \sqrt{-g} \Omega}{\delta s_{ab} \delta s_{cd}} \Bigg|_{s=0} = 
 - \frac{4}{VT}  \frac{\delta^2  W}{\delta \barg_{ab} \delta \barg_{cd}} \Bigg|_{\barg=\eta}~.
 \label{eq:elastic3}
\end{equation}

For a cubic crystal 
\begin{equation}
X^{abcd} = \left(\bar{K} - \frac{2}{3} \bar{\mu} \right)  \delta^{ab} \delta^{cd}   
\ + \ \bar{\mu} 
\, \left( \delta^{ac} \delta^{bd} + \delta^{ad} \delta^{bc} \right) \ + \ 
\alpha \, \delta^{abcd}~, 
\label{eq:elastic2}
\end{equation}
where the term proportional to $\alpha$ is non-zero only if $a=b=c=d$ and represents the anisotropy
in the elastic coefficients. Now we have all the pieces to write the LECs in terms of thermodynamic 
derivatives. For convenience, we define $P=-\frac{1}{3}\langle T^a_a \rangle$, and 
$E=\langle T^{00} \rangle$. 

\subsubsection{The quadratic phonon lagrangian} 

Making the identifications discussed above, we can write the quadratic lagrangian in 
a rather compact form.  Neglecting constant terms and total derivatives, and using integration 
by parts to simplify some terms, we find: 
\begin{eqnarray}
{\cal L}_0  &= &   \frac{1}{2}\bigl[-{F}_{A_0 A_0}\bigr](\partial_0\phi)^2
  - \frac{1}{2}\left[ -\frac{1}{3}\bigl({F}_{A_a A_b}\bigr)\eta_{ab} \right]
  (\partial_i \phi)^2 
  \nonumber \\
&+& \frac{1}{2} \left[ P + E + \frac{m_n^2}{3} \bigl( F_{A_a A_b} \bigr) \eta_{ab} \right]
\dot{\xi}^a \dot{\xi}^a 
\nonumber \\
&+&\left[\frac{1}{3}\bigl({F}_{A_0 g_{ab}}
  + m_n {F}_{A_a A_{b}}\bigr)\eta_{ab} 
  \right] (\partial_c\xi^c)(\partial_0\phi)
  \nonumber \\
&-& \frac{1}{4} \big[ \mu \big] \, \xi^{ab} \, \xi^{ab}   
-\frac{1}{2} \left[K\right] (\partial_c\xi^c)^2
-\frac{1}{2}\alpha\sum_a(\partial_a\xi^a\partial_a\xi^a)
~\label{eq:lowElagrangian_flat}~. 
 \end{eqnarray} 
%
%
%
This form allows us to express the low-energy constants appearing in 
Eq.~\ref{eq:phenomLagrang}  in terms of thermodynamic functions and derivatives as follows, 
\begin{equation}
\begin{split}
\rho&=P + E + \frac{m_n^2}{3} \bigl(F_{A_a A_b}\bigr)\eta_{ab}\\
{K}&=\bar{K}+\frac{1}{3}P \\
{\mu} &= \bar{\mu}-P\\
f_\phi^2&=-F_{A_0A_0}\\
v_\phi^2f_{\phi}^2 &= -\frac{1}{3}F_{A_aA_b}\eta_{ab}\\
g_{\rm mix}&=\frac{1}{\sqrt{\rho}f_\phi}
\bigl[\frac{1}{3}\bigl({F}_{A_0 g_{ab}} + m_n {F}_{A_a A_{b}}\bigr)\eta_{ab}
\bigr]~\label{eq:LECs}\; ~, 
\end{split}
\end{equation}
where
\begin{eqnarray}
\bar{K} &=&  \bigl(\frac{-5}{18}\delta_{abcd} + \frac{1}{6}\delta_{ab}\delta_{cd} + \frac{1}{9}\delta_{ac}\delta_{bd}\bigr) 
X^{abcd}\\
\bar{\mu} &=& \bigl(-\frac{1}{6}\delta_{abcd} + \frac{1}{6}\delta_{ac}\delta_{bd}\bigr) 
X^{abcd}\\
\alpha&=&\bigl(\frac{5}{6}\delta_{abcd} -\frac{1}{6} \delta_{ab}\delta_{cd} -\frac{1}{3} \delta_{ac}\delta_{bd}\bigr) 
X^{abcd}
\;,
\label{eq:mudef}
\end{eqnarray}
with $X^{abcd}$ given in Eqs.~\ref{eq:elastic3},~\ref{eq:elastic2}.
 
The pressure term in the definition of the bulk ($K$) and shear modulus ($\mu$) in
Eq.~\ref{eq:LECs} should not be cause for concern. Its origin can be traced back to the term linear
in $s_{ab}$ in Eq.~\ref{eq:Fexp}, which is present in a system at finite
pressure~\cite{wallace,ray}.  
 
 This dependence of the elastic constants on  the linear term in the Taylor expansion of the free
energy with respect to the strain tensor $s_{ab}$ appears to be counter-intuitive.  However, the
key point here is that the strain tensor  associated  with  a deformation $\xi^a$ has parts both
linear and quadratic in $\partial_b\xi^a$ (Eq.~\ref{eq:strain}).  Using this,  one can show that
the elastic constants are entirely determined by the quadratic terms in the expansion of the free
energy with respect to the displacement field  $\xi^a$ (that appears in the combinations $\xi^{ab}$
and $\partial_a\xi^a$).

\section{Applications~\label{Section:Application}}
\subsection{Neutron star inner crust}
Here we apply the formalism to the inner crust of neutron stars and illustrate the importance of entrainment
and kinetic mixing induced by the neutron-proton interactions. We revise the calculation of the mixing constant $g_{\rm mix}$
(Eq.~\ref{eq:phenomLagrang}) in Ref.~\cite{Aguilera:2008ed} including the effects due to entrainment.
In this earlier work, a somewhat arbitrary distinction was made between neutrons bound in the nuclei
and the neutrons ``outside''. The interaction between the nuclei and unbound neutrons was modeled by
a short-range potential $V(r)=-2\pi a_{nI}~\delta^3(r)/{m_n}$ where $a_{nI}$ was the effective neutron-nucleus
scattering length. Here, using the results of the previous section we show that neither of
these ad hoc assumptions are necessary as the LECs of the effective theory are simply related to
generalized thermodynamic derivatives evaluated in the non-perturbative ground state.  A first
principles calculation of the LECs would require a numerical non-perturbative calculation. Such a
calculation is beyond the scope of this study. In what follows we will use simple estimates based on
earlier calculations to draw some qualitative conclusions about the role of interactions between the
solid and the superfluid in the neutron star crust.   

From the preceding discussions the lowest order effective lagrangian for longitudinal modes with
canonically normalized fields ${\tilde \phi}=f_\phi \phi$ and ${\tilde \xi_i}=\sqrt{\rho}\xi_i$ in
the inner crust can be written as 
\begin{equation}
{\cal L}  =\frac{1}{2} (\partial_0 {\tilde \phi})^2 -\frac{1}{2}v_\phi^2~(\partial_i {\tilde \phi})^2
+  \frac{1}{2}(\partial_0 {\tilde \xi_i})^2 - \frac{1}{2}~v_l ^2 (\partial_i \tilde{\xi}_i)^2  
+ g_{\rm mix}~\partial_0 {\tilde \phi}~ \partial_i {\tilde \xi}_i  \,,
\end{equation}
where the LECs defined in Eq.~\ref{eq:LECs} can be written as 
\begin{eqnarray} 
 v_\phi^2=\frac{n_f}{m_n~f_\phi^2}\,, \quad  v_l^2 = \frac{{K}+(4/3){\mu}}{\rho}\,, {\rm and}\quad g_{\rm mix}=\frac{1}{3} \frac{\bigl({F}_{A_0 g_{ab}} + m_n {F}_{A_a A_b}\bigr)~\eta_{ab}} {\sqrt{-F_{A_0 A_0} \rho}}\,,
\label{eq:NSLECs}
\end{eqnarray}
Here we note that  $(-F_{A_a A_b}\eta_{ab}/3)_f/m_n=(n_n-n_b)/m_n$ where $n_f$ is the density of ``free'' neutrons that participate in superfluid motion, $n_b$ is the number density
of neutrons entrained by the lattice and $n_n$ is the total neutron number density. Further, since
neutrons and ions remain non-relativistic in the neutron star crust, the LECs simplify  
\begin{equation}
\begin{split}
\rho&=E+P  + \frac{m_n^2}{3} \bigl(F_{A_a A_b}\bigr)\eta_{ab}\\
&\xrightarrow{\rm Non-Rel} (n_p+n_b)~m_n  \\
g_{\rm mix}&=\frac{1}{3} \frac{\bigl({F}_{A_0 g_{ab}} + m_n {F}_{A_a A_b}\bigr)\eta_{ab}} 
{f_\phi \sqrt{\rho}}\\
&\xrightarrow{\rm Non-Rel} 
\frac{1}{ f_\phi \sqrt{ \rho}}\bigl[ n_b - n_p~\frac{\partial^2 f}{\partial n_p\partial \mu_n}  \bigr] = \frac{1}{f_\phi \sqrt{ \rho}}\bigl[  n_b - n_p~\frac{\partial n_n}{\partial n_p} \bigr] 
\label{eq:NRLECs}
\end{split}
\end{equation} 
where the hybrid free energy function is 
\begin{equation}
f(\mu_n,n_p)=(\mu_n+m_n) n_n(\mu_n) - E(\mu_n,n_p)\,.
\end{equation}
Here $n_p$ is the proton density, ${{K}}$ and ${{\mu}}$ are respectively the bulk and shear
moduli of the combined system, and we have ignored the small contribution due to the LEC $\alpha$ that encodes the anisotropic contribution. In~\cite{Aguilera:2008ed} a simple estimate of $g_{\rm mix}$ was derived but failed to include the contribution due to entrainment effects. We have
verified that the result in Ref.~\cite{Aguilera:2008ed} can be recovered by setting $n_b=0$ in
Eq.~\ref{eq:NRLECs}. 

In Ref.~\cite{Aguilera:2008ed}, assuming that the effective interaction between the unbound neutrons
and ions is weak, it was found that $f_\phi^2=m_n k_F/\pi$, and $v_\phi^2 = n_f/(m_n f_\phi^2)$,
where $k_F$ and $n_f=k_F^3/3\pi^2$ are the Fermi momentum and number density of unbound neutrons.
The speed of longitudinal lattice vibrations was approximated as the Bohm-Staver sound speed. The
longitudinal sound speed is given by $v_l=\sqrt{{K}_I/\rho}$ where ${K}_I= \rho(\partial (P_{Ie})
/\partial{ \rho})$ is the bulk-modulus of the electron-ion system. To calculate the longitudinal speed
in the Bohm-Staver approximation, the total pressure of the electron-ion system is (well)
approximated by the electron pressure $P_{e}$, and the mass density of lattice is taken to be $\rho =
m_n A$ where $A$ is the number of bound nucleons in the ion.  Interactions between nucleons will
modify these simple estimates quantitatively. Qualitatively, the effect of strong neutron-proton
interactions is the induced mixing between longitudinal lattice phonons and the superfluid modes.
This interaction is characterized by the dimensionless LEC, $g_{\rm mix}$, which in turn depends on
two contributions, one proportional to $n_p(\partial n_n/\partial n_p)$ and the other  proportional
to the entrainment parameter $n_b$.

In the neutron star context, both of these quantities can be calculated using phenomenological
models. The ground state structure which specifies the profiles of nucleons is obtained by solving
the single-particle equations in the Wigner-Seitz (WS) approximation~\cite{Baldo:2005} or more
realistic boundary conditions that reflect the cubic lattice
structure~\cite{Chamel:2005,Newton:2006}. For a given volume of the WS unit cell $V_{\rm WS}$, these
calculations determine the number of bound neutrons ($N_b$), protons ($Z$), and the total number of
neutrons in the cell ($N_{\rm WS}$). They also determine how $\mu_n$ and $\mu_p$ vary with neutron
and proton densities.  The first contribution to $g_{\rm mix}$ is found by noting that $n_p(\partial
n_n/\partial n_p) = n_p f_\phi^2 V_{\rm np}$ where $V_{\rm np} =(\partial \mu_n/\partial n_p)  $  is
the effective interaction between neutrons and protons. The other contribution is related to the
density of bound neutrons and a naive estimate would suggest that $n_b=N_b /V_{\rm WS}$. However, as
discussed earlier in section ~\ref{sec:entrainment} and in Ref.~\cite{Pethick:2010}, the number
density of neutrons that effectively move with the nucleus is defined through the static limit of
the current-current correlation function $\kappa=-F_{A_a A_b}\eta_{ab}/3=(n_n-n_b)/m$. This
correlation function has been computed earlier for neutrons in the background of a static periodic
potential designed to mimic the neutron star crust in Refs.~\cite{Carter:2004pp,Chamel:2005}.  In
these calculations $\kappa=(n_n-n^*)/m^*$ is defined in terms of an ad hoc but convenient quantity
called the effective mass $m^*$ of unbound neutrons, and the average number density of neutrons with
energy greater than zero is denoted by $n^*$.  Thus the  LEC 
\begin{equation} 
n_b=n_n-\frac{m_n}{m^{*}_n}(n_n- n^*)\,, 
\end{equation} 
where 
$n_n=N_{\rm WS}/V_{\rm WS}$ is the average neutron density in the cell. 

We now turn to a simple illustration of how mixing affects the propagation of 
longitudinal modes in the crust. For this purpose it would be ideal to compute the
three LECs ($v_\phi,v_l,g_{\rm mix}$) from a self-consistent underlying microscopic
model using Eq.~\ref{eq:NSLECs}. However, such a calculation is beyond the scope of this work and we
adopt a less rigorous approach where we use the results of  Ref.~\cite{Aguilera:2008ed} for the
velocities of the superfluid and lattice modes in the uncoupled system, and assume that $n_b \gg n_p
(\partial n_n/\partial n_p)$ and $m^*/m \simeq 1$.  Simple estimates support our expectation that
the dominant contribution to $g_{\rm mix}$ is due to $n_b$. In this case, 
\begin{equation}
g_{\rm mix}\simeq v_\phi~\frac{n_b}{\sqrt{(n_b+n_p)n_f}}
\label{eq:gmix_nb}
\end{equation}
Our second assumption $m^*/m \simeq 1$ is likely to be invalid in some regions of the crust
\cite{Chamel:2006}. Nonetheless to simply illustrate the role of mixing we have set $m^*=m$ and plan to
return to a fully self-consistent calculation in future work.   
      
\begin{figure}[t] 
   \centering
   \includegraphics[width=5in]{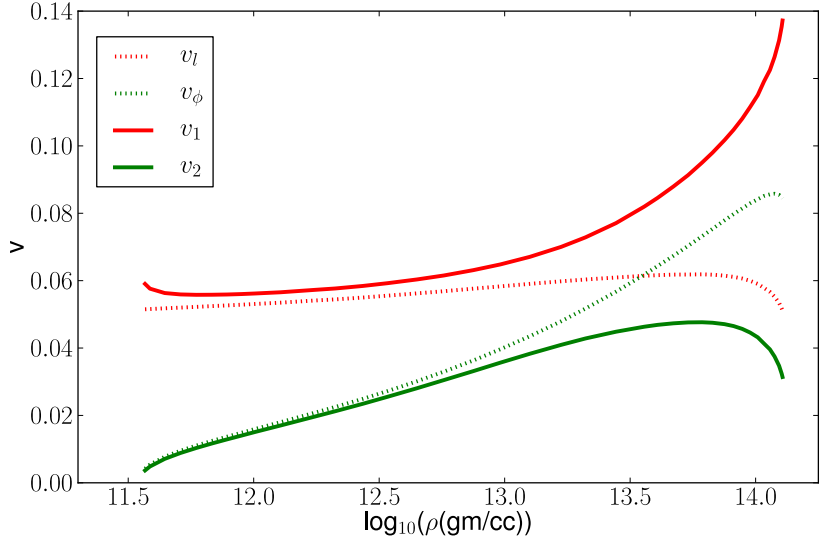} 
   \caption{The velocities of the two eigenmodes. The dotted lines are $v_l$ and $v_\phi$ ignoring
   mixing.}  
   \label{fig:velocities}
\end{figure}

In terms of the canonically normalized fields the kinetic terms in Fourier space has the form,
\begin{equation}
{\calS} =  \frac{1}{2}\sum_k
(\tilde{\varphi}(-k)\;\; \hat{\bfk}.\tilde{\xi}(-k))
\left(\begin{array}{cc}
k_0^2-v_\phi^2\bfk^2&g_{\rm mix}k_0|\bfk|\\
g_{\rm  mix}k_0|\bfk|&k_0^2-v_l^2\bfk^2
\end{array}
\right)
\left(\begin{array}{c}
\tilde{\varphi}(k)\\ 
\hat{\bfk}.\tilde{\xi}(k)
\end{array}
\right)\label{eq:Kinetic}
\end{equation}
The velocities of the two eigenmodes can be obtained by diagonalizing the matrix in Eq.~\ref{eq:Kinetic}.
The results are shown in Fig.~\ref{fig:velocities} where the solid curves incorporate mixing effects due to a finite $g_{\rm mix}$ given by Eq.~\ref{eq:gmix_nb} and the dotted curves show the uncoupled case with $g_{\rm mix}=0$.

In contrast, the speed of the transverse lattice modes are unaffected by mixing and is given by
\begin{equation} 
v_t=\sqrt{\frac{{\mu}}{\rho}} = \sqrt{\frac{{\mu}}{(n_p+n_b)m_n}}\,.
\end{equation}
Here, only entrainment effects play a role in the propagation of transverse lattice phonons, 
as was previously pointed out in Ref.~\cite{Pethick:2010}.

\subsection{Crystalline superfluids or LOFF-like phases}
Other systems of phenomenological interest where this low energy theory applies are the LOFF
phases~\cite{FF:1964,LO:1965}. Here, attractive interactions between two species of fermions leads
to pairing at the Fermi surface but with a pair condensate $\langle \psi_1 \psi_2 \rangle$ which is
spatially inhomogeneous in the ground state. A mismatch in the Fermi-momenta of the two interacting
species in the absence of pairing, disfavors the formation of zero-momentum Cooper pairs and
instead pairs with finite total momentum are favored.
These pairs condense to form a ground state that breaks translation symmetry and can be written as a
sum over plane-waves,
\begin{equation}
\langle\psi_1(x)\psi_2(x)\rangle\sim \Delta(\rr) = \Delta\sum_{\setq{}{a}}
e^{2i\q{}{a}\cdot\rr}\;,
\end{equation}
where $\Delta$ is the gap parameter. The magnitude of the momentum $|{\bf{q}}^a| \simeq \delta k_F$
where $\delta k_F$ is the splitting between the Fermi momenta of the interacting species. The
magnitudes of the momenta and their spatial orientation is determined by minimizing the total free energy and
this set of momenta specifies a crystalline ground state. The LOFF phases can in principle be
realized in cold atomic Fermi gases where a splitting between Fermi levels can be achieved through a
population imbalance~\cite{Combescot:2007} and in dense quark matter where a Fermi level splitting
arises naturally~\cite{Casalbuoni:2004,Alford:2008}. 

In dense quark matter, pairing between different flavors of quarks can play a role in determining
the ground state structure. The relatively large strange quark mass, and charge neutrality induce a
splitting between the Fermi energies of up, down and strange quarks. The expected splitting between
the Fermi energies is $\delta \mu \simeq m_s^2/(8\mu_q)$ where $m_s$ is the strange quark mass and $\mu_q$ is the quark
chemical potential.  At moderate densities, such as those realized in the neutron star core where
$\mu_q \simeq 400$ MeV, this splitting between Fermi energies can favor LOFF phases in quark matter
with spatially varying di-quark condensates with a crystalline structure when $\delta \mu \sim
\Delta_0/\sqrt{2}$~\cite{Alford:2000ze,Bowers:2002xr,Rajagopal:2006}, where $\Delta_0$ is the gap in
the absence of Fermi surface splitting. As this ground state breaks the same symmetries as those
discussed in Section \ref{Section:Formal_development}, these phases are amenable to the same low
energy energy effective theory formulation. 

Several aspects of the low energy theory of crystalline phases in quark matter have already
been described in Ref.~\cite{Casalbuoni:2001gt}. In Ref.~\cite{Mannarelli:2007}, the coefficients of the
``lattice only'' ($\phi=0$ in Eq.~\ref{eq:phenomLagrang}) effective theory were computed
microscopically in a Ginzburg-Landau expansion \cite{Rajagopal:2006}.  This work was
primarily focussed on the shear modes and showed that both the kinetic coefficient, $\rho$, and the
elastic constants ${\mu}$ etc., were of the order $\mu_q^2\Delta^2$ where $\Delta$ is the
pairing gap parameter.  The mixing between the longitudinal lattice
phonon mode and the superfluid mode was mentioned but the relevant mixing coefficient was not
calculated. Using the same techniques, we have estimated the mixing coefficient and we find that in
the regime where LOFF-like phases are favored   
\begin{equation}
g^{\rm LOFF}_{\rm mix}\sim \frac{\Delta}{\delta \mu} \,. 
\end{equation} 
Similarly, the coefficients for the ``superfluid only'' ($\xi^a=0$) sector can also be computed
(some were calculated in~\cite{Gatto:2007ja}). For $\Delta\ll\mu_q$ we expect the coefficient
$f_\phi^2\sim \mu_q^2$, corresponding to the density of states near the Fermi surface for a
relativistic system.  Our simple estimates here show that strong mixing between the superfluid
and the longitudinal mode can be realized, with important implications for hydrodynamic oscillations
both in the context of dense quark matter and trapped imbalanced Fermi gases, where LOFF phases may
also be potentially realized. Definitive results require a rigorous derivation of the low-energy
constants and such a calculation is being pursued and will be reported elsewhere. 

\section{Conclusions}
\label{Section:conclusion}

We have studied  a low energy effective theory describing phases of matter that simultaneously break
translational symmetry and a number conservation symmetry. $U(1)$ phase invariance and general
coordinate invariance restrict the combinations of terms that can appear in the effective
lagrangian. We have shown that the lowest order lagrangian (featuring equal number of derivatives
and Goldstone fields)  is determined  by the derivatives of the thermodynamic pressure with respect to
the external fields such as the chemical potential. While this was known in the case of one
superfluid system~\cite{Son:2002zn,Son:2005rv}, here we have  provided a different proof for
superfluids and we have generalized it to the mixed system. The two main results of this paper,
Eqs.~\ref{eq:W2derivatives} and~\ref{eq:lowElagrangian_flat}, provide a useful framework for
computing the low energy dynamics.   

Our thermodynamic matching relates the LECs to thermodynamic derivatives  of the free energy with
respect to external fields (chemical potential, vector potential, background metric).  We have also
pointed out the relation between LECs and  correlators of the $U(1)$ current and the stress tensor
at small momenta.  Both approaches might be pursued in future non-perturbative calculations using
many-body techniques such as Skyrme Hatree-Fock and Quantum Monte Carlo.  

As a concrete example of  phenomenological interest, we have considered matter in the inner crust of
a neutron star,  updating a previous estimate of the parameter characterizing the kinetic mixing of
superfluid and lattice phonons. We also discussed briefly how this formulation would apply to the
crystalline superfluids or LOFF-like phases and highlighted the role of mixing between the modes in
these systems.

Finally, we note  that the formalism that we have set up here can be applied to study
the low-energy dynamics of other  physical systems with several spontaneously broken symmetries,
such as a system composed of two superfluid species.

\vspace{.5cm}
{\bf Acknowledgements}  We  thank Tanmoy Bhattacharya,  Nicolas Chamel, Michael Forbes, Michael
Graesser, Emil Mottola, Chris Pethick, and  Dam Son for useful discussions at various stages of this work. We
thank Krishna Rajagopal and Massimo Mannarelli for comments and suggestions on the manuscript. This
work was supported by grants from the department of energy DE-AC52-06NA25396 (LANL), and the DOE
topical collaboration to study of Neutrinos and nucleosynthesis in hot and dense matter.

\newpage

\appendix
\section{$\Omega[\barA^n,\barA^p,\barg]$ and energy density of deformed states}
\label{sec:app1}

In this appendix we show that  the  energy density   $\Omega [\barA^n,\barA^p,\barg] = 
-W [\barA^n,\barA^p,\barg] / (V T)$  calculated using the path
integral (Eqs.~\ref{eq:Zmixed} and \ref{eq:barOmega}) admits a simple physical interpretation. 
It is the expectation value per unit volume 
${\cal E} [\zeta_g]  = 
 \langle \Omega_g |  \hat{H}_{\barA^n,\barA^g,\barg=\eta} | \Omega_g \rangle / V$  
of the flat-space Hamiltonian 
in the state $| \Omega_g \rangle$  that minimizes 
${\cal E} [\zeta_g]$ 
subject to the constraint $\langle \Omega_g | \xi^a (x) | \Omega_g \rangle =  \zeta_g^a (x)$,
with  $\zeta_g^a(x)$  satisfying  $\barg_{ab} = \eta_{ab} - 2 s_{ab} (\zeta)$ (see Eq.~\ref{eq:strain}).
In other words  $\Omega [\barA^n,\barA^p,\barg]$ 
 is the energy density in the lowest energy state subject to the 
 ``deformation condition" $\langle \Omega_g | \xi^a (x) | \Omega_g \rangle =  \zeta_g^a (x)$.  
It is precisely in this sense one should think of the metric $\barg_{ab}$ as determining the shape of the 
system. 
To avoid notational clutter, we will  focus here on the case of a pure solid system and 
neglect the dependence on the external fields $\barA^n$ and $\barA^p$. 
The derivation involves several steps, which we summarize below. 

\begin{itemize}

\item First,  let us evaluate the partition function  
in the presence of a space-time independent background metric $\bar g_{\mu \nu}$  of form Eq.~\ref{eq:g0}  
by the saddle point method.    
The classical solution  that minimizes the Euclidean action 
and is well behaved at $|x|\rightarrow\infty$ is given by $\xi^a=0$.
So we have: 
\begin{eqnarray}
Z[\barg] =  e^{i W[\barg]} &=&  
{\rm Exp} \left\{ i V T  \, {\cal  L}_0   \Big( H^{a b} (g = \barg, \xi = 0) \Big)
\right \} 
\end{eqnarray}

\item Since we are working with  a diffeomorphism invariant theory, 
we can  obtain the same result for the free energy in a 
different coordinate system. 
Let us use this freedom to  switch from coordinates 
$(x^a, \barg_{ab})$  to the ``flat" coordinates   
$(\tilde{x}^a, \eta_{ab})$. \footnote{The flat coordinates  
$(\tilde{x}^a, \eta_{ab})$ play  a somewhat special role:  
the configuration $\tilde{\xi}^a=0$ corresponds to the 
equilibrium configuration in absence of external fields. 
In this state the body-fixed coordinate 
$z^a = \tilde{x}^a - \tilde{\xi}^a$   are flat (coincide with the laboratory 
coordinates).  Deformations from equilibrium $\tilde{\xi} \neq 0$ induce a 
non-euclidean metric in the body-fixed coordinates.  
}
The appropriate variable transformation can be found by noting that 
$H^{ab}$ is a scalar density. 
This results in  $x^a (\tilde{x}) = \tilde{x}^a  - \xi^a_g (\tilde{x})$,  with the field $\xi_g$ 
determined by the condition 
$H^{ab} (g=\barg, \xi=0) = H^{ab} (g=\eta, \xi = \xi_g)$, which explicitly reads
\beq
\barg^{ab}  = \eta^{ab}  - \Big( 
\frac{\partial  \xi_g^b
}{\partial{\tilde{x}_a}}    + 
\frac{ \partial  \xi_g^a}{\partial \tilde{x}_b}  - \eta^{ij} 
\frac{ \partial \xi^a_g }{\partial \tilde{x}_i}  \, 
\frac{ \partial \xi^b_g }{\partial \tilde{x}_j}  \, 
\Big)~. 
\label{eq:gabvsxi2}
\eeq
Eq.~\ref{eq:gabvsxi2}  defines 
$\xi_g^a$ up to rigid rotations and translations. 
For constant $\barg_{ab}$ the solution has the form 
$\xi_g^a(\tilde{x}) = K^a_b \tilde{x}^b + c^a$ where the elements of $K^a_b$ and $c^a$ are constant  
\footnote{Note however, that one needs to avoid ``large diffeomorphisms'', which are
not well behaved at $|x|\rightarrow\infty$. 
Proper behavior at infinity can be ensured by multiplying the transformation 
by appropriate convergence factors that decay to
zero at $|x|\rightarrow\infty$ faster than any polynomial.}.
Equivalently  the inverse change of variables reads  $\tilde{x}^a (x) = x^a  + \zeta^a_g (x)$, 
with $\zeta^a_g (x)  = \xi^a_g (\tilde{x})  + O(\xi_g^2)$, and one has  
$\barg_{ab} =  \eta_{ab}  - 2  s_{ab} (\zeta)$,  with the strain  $s_{ab} (\zeta)$ 
given in Eq.~\ref{eq:strain}.\\
In summary,  as a consequence of general coordinate invariance one has: 
\begin{eqnarray}
Z[\barg] =  e^{i W[\barg]} &=&  
{\rm Exp} \left\{ i V T  \, {\cal  L}_0   \Big( H^{a b} (g = \eta, \xi = \zeta_g ) \Big)
\right \}  ~,
\label{eq:proof3}
\end{eqnarray}
with a time-independent field configuration  $\zeta_g (\vec{x})$ determined by Eqs.~\ref{eq:strain} 
or alternatively \ref{eq:gabvsxi2}.

\item Next we note that the exponent on the RHS of  Eq.~\ref{eq:proof3}  
is the {\it  flat-space} action evaluated at the  field $\xi= \zeta_g$. 
Moreover, to leading order in the loop expansion (and low-energy expansion) 
the action coincides with the quantum effective action $\Gamma [\zeta_g] = {\cal S}_{\rm eff} [\zeta_g]$. 
But the quantum effective action $\Gamma [\zeta_g]$  admits an energy interpretation 
\cite{Symanzik:1969ek,Coleman:1974hr,Weinberg:1996kr}:
for time-independent field configurations $\zeta_g (\vec{x})$,   one has that   
$\Gamma [\zeta_g]/T   =   -   \,   \langle \Omega_g |  \hat{H}_{\barA^n,\barA^g,\barg=\eta} | \Omega_g \rangle$
where  $ |\Omega_g \rangle$  is the state that minimizes the expectation value of the 
Hamiltonian under the constraint 
$\langle \Omega_g | \hat{\xi}^a (x) | \Omega_g \rangle =  \zeta_g^a (\vec{x})$~.   
In equations, the above chain of reasoning reads
\bea
W[\barg] & \equiv & - V T \, \Omega[\barg] 
\\
W[\barg] &=& {\cal S}_{\rm eff} [\zeta_g] = 
\Gamma [\zeta_g] =  
- T   \,   \langle \Omega_g |  \hat{H}_{\barA^n,\barA^g,\barg=\eta} | \Omega_g \rangle \\
& \equiv & - V T \,  {\cal E} [\zeta_g]~, 
\eea
thus proving that $\Omega [\barg] = {\cal E}  [\zeta_g]$, 
with   $\zeta_g$  related to $\barg_{ab}$ by 
$\bar{g}_{ab} = \eta_{ab} - 2 s_{ab} (\zeta_g)$. 

\end{itemize} 

\bibliographystyle{unsrt}
\bibliography{phonons} 
\end{document}